\documentclass[english,11pt,aps,prd,a4paper,preprintnumbers,floatfix,nofootinbib,showpacs,superscriptaddress, notitlepage]{revtex4-1} 

\usepackage[utf8x]{inputenc} 
\usepackage{booktabs}
\usepackage{color}
\usepackage{graphicx}
\usepackage{comment}

\usepackage{multirow}
\usepackage{amsmath}
\usepackage{amssymb}
\usepackage[colorlinks=true,citecolor=darkred,urlcolor=darkred, pdfborder={0 0 0}]{hyperref}
\usepackage[normalem]{ulem}
\usepackage{soul}
\definecolor{darkred}{rgb}{0.6,0,0}
\definecolor{drkgrn}{RGB}{0, 51, 0}
\definecolor{gray}{RGB}{128, 128, 128}
\usepackage{soul}

\def\beq{\begin{equation}}
\def\eeq{\end{equation}}

\def\u1p{$U(1)^{\prime}$~}

\def\FO{\textit{freeze-out}~}
\def\FI{\textit{freeze-in}~}
\setstcolor{red} 
\newcommand{\bea}{\begin{eqnarray}}
\newcommand{\eea}{\end{eqnarray}}

\newcommand{\AddrINFN}{Istituto  Nazionale  di  Fisica  Nucleare,  Sezione  di  Bari,  Via  Orabona  4,  70126  Bari,  Italy}
\newcommand{\AddrAlabama}{Department of Physics and Astronomy,  University of Alabama, Tuscaloosa,  AL 35487, USA}
\newcommand{\AddrDelaware}{Bartol Research Institute,  Department of Physics and Astronomy,  University of Delaware,  Newark,  DE 19716, USA}
\newcommand{\AddrWColl}{Physics Department,  Washington College,  Chestertown,  MD 21620, USA}

\begin{document}

\title{Light $Z^\prime$ and Dirac fermion dark matter in the $B-L$ model}

\author{Newton Nath}\email{newton.nath@ba.infn.it}\affiliation{\AddrINFN}
\author{Nobuchika Okada}\email{okadan@ua.edu}\affiliation{\AddrAlabama}
\author{Satomi Okada}\email{satomi.okada@ua.edu }\affiliation{\AddrAlabama}
\author{Digesh Raut}\email{draut@udel.edu}\affiliation{\AddrDelaware} \affiliation{\AddrWColl}
\author{Qaisar Shafi}\email{qshafi@udel.edu}\affiliation{\AddrDelaware}

\begin{abstract}
{\noindent 
We consider a $U(1)_{B-L}$ model with a $Z^\prime$ portal Dirac fermion dark matter (DM) $\chi$ of low mass which couples very weakly to the $B - L$ gauge boson $Z^\prime$. 
An arbitrary $B-L$ charge $Q\neq \pm1, \pm 3$ of the DM $\chi$ ensures its stability. 
Motivated by the sensitivity reach of forthcoming ``Lifetime Frontier" experiments, we focus on the $Z^\prime$ mass, $m_{Z^\prime}$, in the  sub-GeV to few GeV range. 
To evaluate the DM relic abundance, we examine both the \FO and \FI DM scenarios. 
For the \FO scenario, we show that the observed DM abundance is reproduced near the $Z^\prime$ resonance, $m_\chi \simeq m_{Z^\prime}/2$, where $m_\chi$ is the DM mass. 
For the \FI scenario, we focus on $m_\chi \ll m_{Z^\prime}$. 
We show that for a fixed value of $m_{Z^\prime}$, $g_{BL}$ values roughly scale as $1/Q$ to reproduce the observed DM abundance. 
For various $Q$ values in the range between $10^{-6}$   and $10^2$, we show that the gauge coupling values $g_{BL}$ needed to reproduce the observed DM abundance lie in the search reach of future planned and/or proposed experiments such as FASER, Belle-II, LDMX, and SHiP. 
In the \FI case, the $Q$ values to realize observable $g_{BL}$ values are found to be much smaller than that in the \FO case. }
\end{abstract}

\maketitle

\section{Introduction}
Despite its remarkable success in explaining a large collection of phenomena observed in nature, the Standard Model (SM) of particle physics has several serious shortcomings.  
A viable candidate for dark matter (DM) particle is missing in the SM, presumably an electrically neutral non-baryonic particle that accounts for around 85\% of the observed matter in the Universe ~\cite{Bertone:2004pz, Feng:2010gw}. 
The SM also fails to account for the observed non-zero neutrino masses inferred from solar and atmospheric neutrino oscillation experiments~\cite{SNO:2001kpb,Super-Kamiokande:1998kpq}. 

The $B - L$ (Baryon number minus Lepton number) model~\cite{Davidson:1978pm, Mohapatra:1980qe, Marshak:1979fm, Wetterich:1981bx, Masiero:1982fi} is a simple extension of the SM that has received much attention for successfully explaining the non-zero neutrino masses. 
It is obtained by gauging the accidental global $B-L$ symmetry in the SM. 
The model requires three Majorana right-handed neutrinos (RHNs), one per SM generation, to cancel the $B-L$ related anomalies. 
In the presence of the RHNs, the type-I seesaw mechanism \cite{Minkowski:1977sc, Yanagida:1979as, GellMann:1980vs, PhysRevD.20.2986, Mohapatra:1979ia} to explain the tiny masses of the observed neutrinos is automatically implemented. 
This model can accommodate a variety of DM scenarios. 
For instance, one of the RHNs can serve as a DM candidate if it carries either a $Z_2$ odd charge \cite{Burell:2011wh, Okada:2012sg, Basak:2013cga, Okada:2016gsh, Okada:2017dqs, Okada:2018ktp, Okada:2020cue, Das:2021nqj} or a non-conventional $B-L$ charge \cite{Okada:2018tgy}. 
Another possibility is a Dirac fermion DM $\chi$ ~\cite{Fayet:2004bw, FileviezPerez:2019cyn, Mohapatra:2019ysk, delaVega:2021wpx} with $B-L$ charge $Q\neq \pm1, \pm 3$, which is essential to ensure its stability. 
In all of these neutrino$ - $DM scenarios, the interaction between DM and the SM particles is primarily mediated by the $B-L$ gauge boson $Z^\prime$. 
Hence, these scenarios predict a one-to-one correspondence between the collider searches for $Z^\prime$ and DM physics.

In this work we consider a Dirac fermion DM scenario with arbitrary charge $Q$ and $m_\chi \leq m_{Z^\prime}$, where $ m_\chi $ and $ m_{Z^\prime} $ respectively stand for the DM and $Z^\prime$ boson masses. 
The case with $m_\chi > m_{Z^\prime}$ was examined in Ref.~\cite{Mohapatra:2019ysk}.  
Motivated by the sensitivity reach of the forthcoming Lifetime Frontier experiments, we focus on low mass DM in the range of sub-GeV to a few GeV that couples weakly to the $Z^\prime$ gauge boson. 
We consider both the \FO and  \FI scenarios to evaluate the observed  DM  relic abundance. 
In the \FO scenario, the DM is initially in thermal equilibrium. At late times, when the expansion rate of the Universe dominates over the DM annihilation rate, the DM decouples from the thermal plasma and its abundance saturates to a constant value.  
We find that the correct relic abundance is realized for $m_\chi \simeq m_{Z^\prime}/2$.  
In the \FI scenario, the DM abundance builds up from an essentially zero initial abundance and saturates to a constant value once the expansion rate of the Universe dominates and suppresses its production. 
In this case, the DM never gets into thermal equilibrium and we find that the correct relic abundance is realized for $m_\chi \ll m_{Z^\prime}$. 
 
Various current and future planned/proposed experiments can search for the low mass and small gauge coupling of $Z^\prime$ predicted by our model. 
Seminal particle physics experiments such as BABAR \cite{Lees:2014xha}, LHCb~\cite{Aaij:2019bvg}, COHERENT~\cite{COHERENT:2019kwz} and  several electron and proton beam dump experiments can narrow down the allowed parameter space. 
We also examine constraints from DM direct detection experiments such as Dark-Side-50 \cite{DarkSide:2018bpj}, PandaX-II \cite{PandaX-II:2017hlx}, LUX~\cite{LUX:2019npm}, and the Xenon1T \cite{XENON:2020gfr}, which set tight constraints on the DM - nucleus cross-section. 
Finally, the allowed parameter regions that can be probed by the “Lifetime Frontier” experiments such as FASER and FASER2 ~\cite{FASER:2018eoc,FASER:2019aik}, Belle II, SHiP ~\cite{Alekhin:2015byh} and LDMX~\cite{Berlin:2018bsc} have also been investigated.

%
%
%
%

The paper is organized as follows. Sec. \ref{sec:Framework} is dedicated to a brief description of our theoretical framework, and the dark matter relic density is discussed in Sec. \ref{sec:DMrelic}. We evaluate the DM relic density for the \FO and \FI mechanisms in Sec. \ref{sec:FreezeOUT} and Sec. \ref{sec:FreezeIN}, respectively. A summary of our work and concluding remarks are presented in Sec. \ref{sec:Conclusion}.

\section{The Framework: Dirac  dark matter}\label{sec:Framework}
We consider a gauged $U(1)_{B-L}$ extension of the SM and 
the particle content of the model along with $B-L$ charges  are listed in Table \ref{tab:modelcharges}. 
The SM particles all carry a non-zero $B-L$ charge except for the SM Higgs doublet $H$, whereas the new particles are SM singlets carrying only the $B-L$ charge.  
The three right-handed neutrinos (RHNs) carry $B-L$ charge $-1$, which ensures the cancellation of all the $B-L$ related anomalies.
The vector-like Dirac fermion $\chi$ (DM) and the $B-L$ Higgs $\phi$ carry $B-L$ charges $Q$ and $2$, respectively. 
The arbitrary charge $Q$ is chosen to ensure the stability of the DM.

The scalar potential involving $H$ and $\phi$ is given by
\begin{equation}
V = \lambda_H \left (H^\dagger H - \frac{v_H^2}{2}\right)^2 +\lambda_\phi \left (\phi^\dagger \phi - \frac{v_{BL}^2}{2}\right)^2 + \lambda_{H\phi} \left (H^\dagger H - \frac{v_H^2}{2}\right) \left (\phi^\dagger \phi - \frac{v_{BL}^2}{2}\right) \;.  
\end{equation}
Here, $v_H$ and $v_{BL}$ are the vacuum expectation values (VEVs) of $H$ and $\phi$, respectively, and are expressed as
\begin{eqnarray}\label{eq:HiggsVEV}
\langle H \rangle
 &=&
\frac{1}{\sqrt{2}}
\left(\begin{array}{c}
 v_H\\
 0\end{array} \right), ~~~~
\langle \phi \rangle = \frac{v_{BL}}{\sqrt{2}} \;.
\end{eqnarray}

\begin{table}[t!]
  \centering
  $ \begin{array}{|c|c|c|c|c|c|c|c|c|c|} \hline
    \text{Symmetry/Fields} & Q^i_L & u^i_R & d^i_R & L^i_L &e^i_R & N^i_R &H & \phi & \chi\\ 
    \hline
    U(1)_{B-L} & 1/3 & 1/3 & 1/3 & -1 & -1 & -1  & 0 & 2 & Q \\ \hline
  \end{array} $
  \caption{\footnotesize Particles content and $U(1)_{B-L}$ charges of the model, where $ i = 1, 2, 3$ represent the three generation indices. Here, besides three-right handed neutrinos, a scalar singlet $ \phi$ and a vector-like Dirac DM particle $\chi$ are also present.
  }
  \label{tab:modelcharges}
\end{table}

The Yukawa Lagrangian invariant under ${\rm SM}\otimes U(1)_{B-L}$ for neutrinos is expressed as 
\begin{eqnarray}
 - {\cal{L}}_{\textrm{Y}} \supset \ \sum^3_{i, j = 1} Y^{ij}_D \overline{L^i_L} H N^j_R
    + \frac{1}{2} \sum^3_{i = 1} Y^i_N \phi \overline{N^{iC}_R} N^i_R + h.c.\;,
 \label{Yukawa_BL}
\end{eqnarray}
where $Y_D^{ij}$ and $Y_N^i$ are Dirac and Majorana Yukawa coupling constants, respectively. 
Through these Yukawa interactions, the Dirac and Majorana neutrinos masses are expressed as follows:
\begin{eqnarray}
 m_D^{ij} &=& \frac{Y_D^{ij}}{\sqrt{2}} v_H,  ~~ ~~  
 M_R^i   = \frac{Y_N^i}{\sqrt{2}} v_{BL}\;.
 \end{eqnarray}

The relevant terms in the Lagrangian involving $Z^\prime$ and the DM $\chi$ are given by  
 \begin{equation}
  \mathcal{L} \supset i g_{BL} Q \left(\overline{\chi}\gamma^\mu \chi \right) Z_\mu^\prime + i g_{BL} Q_f \left(\overline{f}\gamma^\mu f \right) Z_\mu^\prime + \frac{1}{2}m_{Z^\prime}^2 Z_\mu^\prime {Z^\prime}^\mu + m_{\chi} \overline{\chi} \chi+h.c. \;,
\end{equation}
where $Q_f$ represent the $B-L$ charge of the fermion $f$, $m_\chi$ is the DM mass, and $m_{Z^\prime} = 2 g_{BL}v_{BL}$ is $Z^\prime$ boson mass obtained from the $B-L$ symmetry breaking. 

\section{DM relic density}\label{sec:DMrelic}
The relic density of the DM particle can be evaluated by solving Boltzmann equation, 
\bea 
 \frac{dY}{dx}
 = - \frac{\langle \sigma v \rangle}{x^2}\frac{s (m_{\chi})}{H(m_{\chi})} \left( Y^2-Y_{EQ}^2 \right) \;,
\label{eq:BE}
\eea  
where, $Y = n/s$ is the yield of the DM particle represented as a ratio of the DM number density ($n$) and entropy density ($s$), $x = m_{\chi}/T$ is a dimensionless parameter with $T$ being the temperature of the Universe, $Y_{EQ}$ is the yield of the DM particle in thermal equilibrium, $H$ is the Hubble parameter, and $\langle\sigma v\rangle$ is the thermally averaged cross-section for a pair annihilation of DM particles times relative velocity $v$. 
The quantities $s$, $H$, $Y_{EQ}$ and $\langle\sigma v\rangle$ are expressed as 
\bea
s &=& \frac{2 \pi^2}{45} g_\star m_{\chi}^3 \;, \nonumber 
\\
 H &=& \sqrt{\frac{\pi^2}{90} g_\star} \frac{m_{\chi}^2}{M_P} \;, \nonumber 
 \\ 
 Y_{EQ}&=& \frac{g_{\chi}}{2 \pi^2} \frac{x^2 m_{\chi}^3}{s(m_{\chi})} K_2(x) \;,  \nonumber 
 \\
 \langle \sigma v_{rel} \rangle &=& \frac{g_{BL}^2}{64 \pi^4}
 \left(\frac{m_{\chi}}{x}\right) \frac{1}{n_{EQ}^{2}}
 \int_{4 m_{\chi}^2}^\infty ds \; 2 (s- 4 m_{\chi}^2) \sigma(s) \sqrt{s} K_1 \left(\frac{x \sqrt{s}}{m_{\chi}}\right)\;.
 \label{eq:DMeqns}
\eea
Here, $g_\star$ and $g_{\chi} (= 4)$ are the thermal degrees of freedom of the plasma and DM particle, respectively, $K_\alpha$ is modified Bessel functions of second kind, respectively, and $n_{EQ}=s(m_{\chi}) Y_{EQ}/x^3$ is the equilibrium number density of the DM particle. 
The relic density of the DM particle at the present time $(x\to \infty)$ is given by 
\bea 
 \Omega_{DM} h^2 =\frac{m_{\chi} s_0 Y(x\to\infty)} {\rho_c/h^2}, 
\label{eq:relicabundance}  
\eea 
where $\rho_c/h^2 =1.05 \times 10^{-5}$ GeV/cm$^3$ is the critical density and $s_0 = 2890$ cm$^{-3}$ is the entropy density of the present Universe. 
From the Planck satellite measurements, $\Omega_{DM} h^2 = 0.1200 \pm 0.0012$ \cite{Aghanim:2018eyx}.

For both freeze-in and freeze-out scenarios, we find that the $Z^\prime$ boson resonance process $\chi \chi \leftrightarrow f_{i} {\bar f}_{i} $, where $f_{i} $ denote SM fermions, dominate the DM annihilation/creation.  
The cross-section for this process is given by
\bea
\sigma (\chi \chi \to \Sigma_{i }f_i{\bar f}_i ) =12 \pi \frac{s}{s-4m_\chi^2} \frac{1}{\left(s-m_{Z^\prime}^2\right)^2+ m_{Z^\prime}^2 \Gamma_{Z^\prime}^2 } \Gamma_{\chi} (s) \times \Big(\Sigma_f \Gamma_{f_i} (s) \Big)\;,  
\eea
for $s \geq 4 m_{\chi}^2$. 
The functions $\Gamma_{f} (s)$ and $\Gamma_{\chi} (s)$ for $s= m_{Z^\prime}^2$ correspond to the partial decay widths of $Z^\prime$ boson at rest and are expressed as 
\bea
 \Gamma_{f} (s) &=& N_f \frac{\left({g_{BL}Q_f}\right)^2}{12\pi} \sqrt {s}  \left(1+\frac{2m_f^2}{s}\right)  \sqrt{1- \frac{4m_f^2}{s}}\;, 
\nonumber \\
  \Gamma_{\chi} (s) &=& \frac{\left({g_{BL}Q_\chi}\right)^2}{12\pi} \sqrt {s}  \left(1+\frac{2m_\chi^2}{s}\right)  \sqrt{1- \frac{4m_\chi^2}{s}}\;, 
\eea 
respectively, and the total decay width of $Z^\prime$ boson $\Gamma_{Z^\prime} = \Sigma_f \Gamma_f (s = m_{Z^\prime}^2) + \Gamma_{DM} (s = m_{Z^\prime}^2)$. 
The values of $N_f =1 (1/2)$ and $Q_f= -1 (-1)$ for the SM leptons $f =  {e,\mu,\tau} \; (\nu_{e,\mu,\tau})$ and for the SM quarks $N_f =3$ and $Q_f= 1/3$. 
Since the gauge coupling values of interest $g_{BL} \ll 1$, we employ the narrow width approximation,
\bea
 \int ds \frac{1}{\left(s-m_{Z^\prime}^2\right)^2+ m_{Z^\prime}^2 \Gamma_{Z^\prime}^2 } \simeq \int ds \frac{\pi}{m_{Z^\prime} \Gamma_{Z^\prime}}\delta({s-m_{Z^\prime}^2})\;. 
\eea
Using this the thermally averaged cross section in Eq.~(\ref{eq:DMeqns}) for the $Z^\prime$ boson resonance process can be expressed as  
\bea
\langle\sigma v_{rel} \rangle \simeq \frac{3\pi^2}{2} \frac{m_{Z^\prime}^2}{m_{\chi}^5}  \frac{x \times K_1 \left(\frac{x \times m_{Z^\prime}}{m_\chi}\right)}{\Big(K_2(x)\Big)^2} \left(\frac{\Gamma_{\chi} (s) \times \Sigma_f \Gamma_{f_i} (s) }{\Gamma_{Z^\prime}}\right)\Bigg|_{s= m_{Z^\prime}^2}\;.  
\label{eq:TAvgCs1}
\eea
For $m_\chi \sim m_{Z^\prime}/2$, we expect an enhancement of the DM pair annihilation cross section through the ${Z^\prime}$ boson resonance. 
So, we parameterize $m_\chi = ( m_{Z^\prime}/2) (1-\delta)$, where $0 < \delta \ll 1$. 
Since $ \Gamma_{\chi} (s = m_{Z^\prime}^2) \ll \Sigma_f \Gamma_f (s = m_{Z^\prime}^2)$ for $\delta \ll1$, Eq.~(\ref{eq:TAvgCs1}) simplifies to 
\bea
\langle\sigma v_{rel} \rangle \simeq \frac{3\pi^2}{2} \frac{m_{Z^\prime}^2}{m_{\chi}^5}  \frac{x \times K_1 \left(\frac{x \times m_{Z^\prime}}{m_\chi}\right)}{\Big(K_2(x)\Big)^2} \times \Gamma_{\chi} (s = m_{Z^\prime}^2) \;,   
\label{eq:TAvgCs}
\eea
which is independent of final state SM particles. 


\subsection{Freeze-out Dark Matter }\label{sec:FreezeOUT}

In the \FO scenario, the DM interaction is strong enough to maintain  thermal equilibrium with the thermal plasma at  early times. 
At late times this is not the case which effectively switches off the DM interaction and the latter decouples from the plasma. 
Hence, the DM abundance effectively remains unchanged after the decoupling.

\begin{figure}[t]
\begin{center}
\includegraphics[scale=1.2]{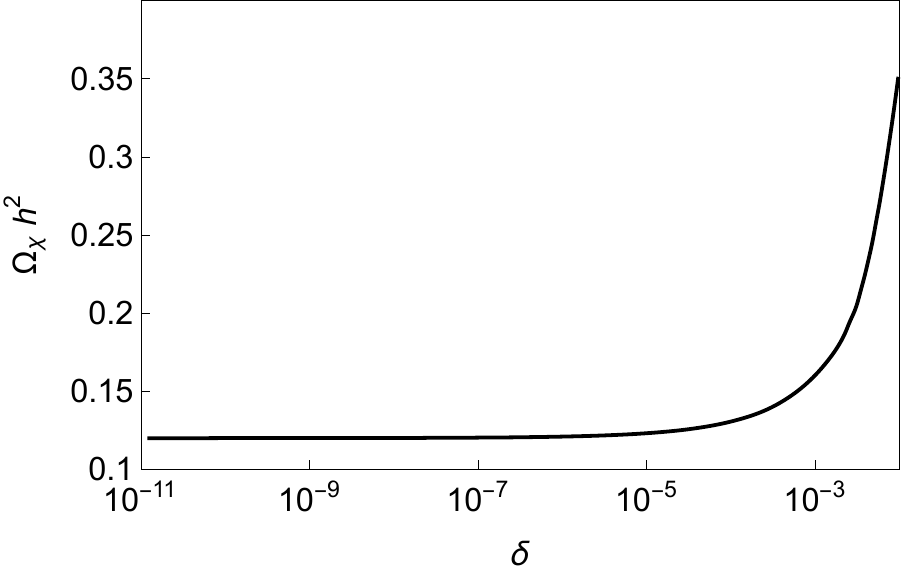}    
\end{center}
\caption{
\footnotesize Relic abundance as a function of $\delta$ for $g_{BL} = 3.46 \times 10^{-5}$, $m_{Z^\prime} = 10$ GeV and $Q_{\chi} = 1$}
\label{fig:omegadelta}
\end{figure}

To obtain the DM yield after decoupling, we solve the Boltzmann equation  given by Eq.~(\ref{eq:BE}) with the initial condition $Y(x\ll1) = Y_{EQ}$. 
The free parameters in numerically solving the Boltzmann equation are $Q$, $g_{BL}$, $m_{Z^\prime}$, and $m_\chi$. 
For small gauge coupling values, the total thermally averaged cross section in Eq.~(\ref{eq:TAvgCs}) is dominated by the contribution from the region very close to the resonance $m_\chi \simeq m_{Z^\prime}/2$ \cite{Mohapatra:2019ysk} which corresponds to $\delta \ll1$. 
We find that the relic density of the DM approaches a constant value in the $\delta \to 0$ limit. 
Our result is shown in Fig.~\ref{fig:omegadelta} for benchmark values $g_{BL} = 3.46 \times 10^{-5}$, $m_{Z^\prime} = 10$ GeV and $Q_{\chi} = 1$. 
We have verified that the asymptotic behavior of the relic density for $\delta \ll1$ is also valid for different choice of input parameters. 
For $Q_\chi=1/3$, the resultant DM relic density is consistent with that in Ref.~\cite{delaVega:2021wpx}, where the authors obtained the result by employing  micrOMEGAs~\cite{Belanger:2018ccd}.

\begin{figure}[t]
\begin{center}
\includegraphics[scale=0.6]{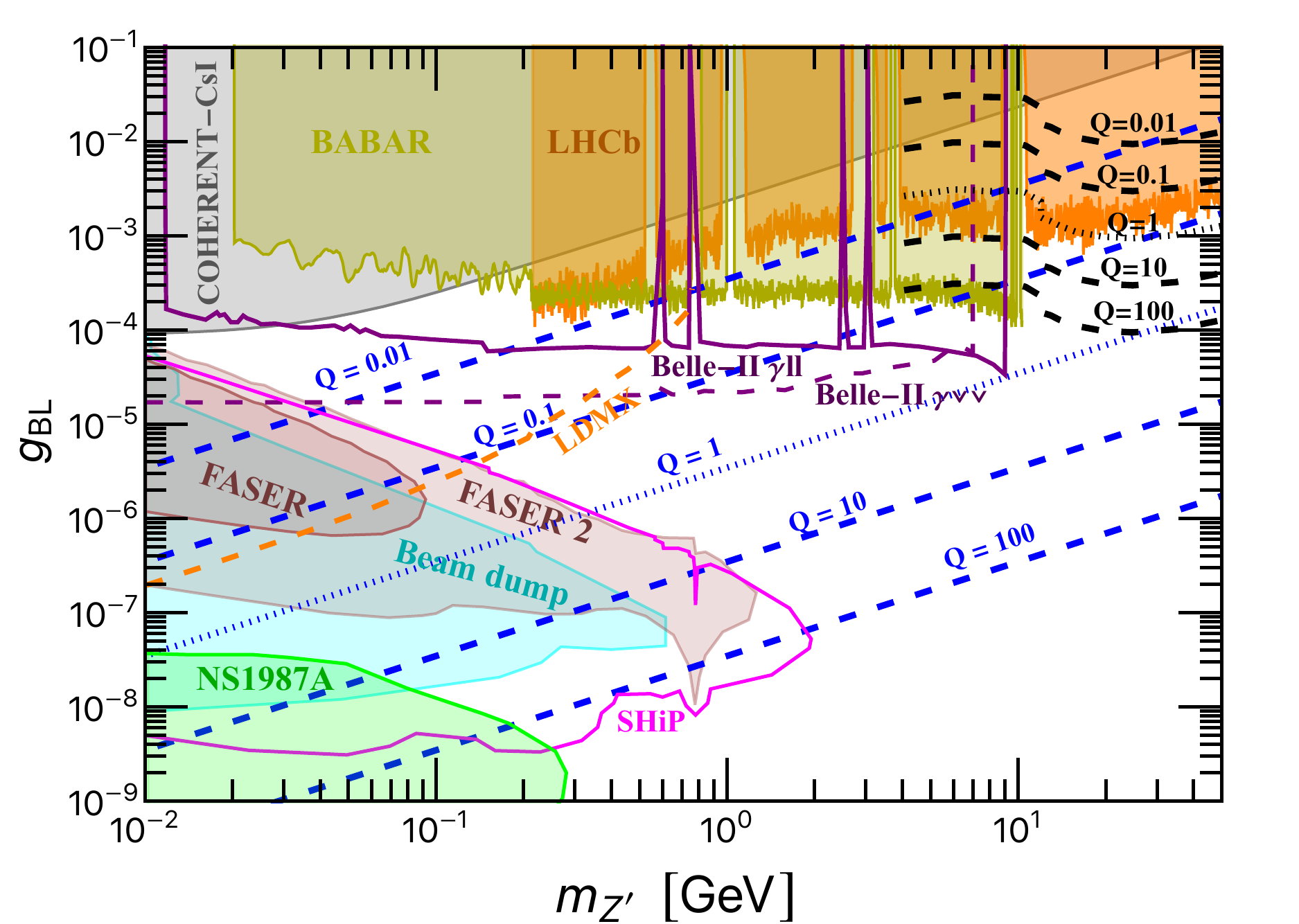}    
\end{center}
\caption{
\footnotesize Exclusion regions in the $(M_{Z^\prime} , g_{BL})$ plane for the \FO DM scenario. 
The diagonal blue lines from top to bottom correspond to $Q = 0.01, 0.1, 1, 10$, $100$, respectively. Along these lines, $100\%$ of the observed DM is reproduced. 
For $m_{Z^\prime} > 4$ GeV, the curved black lines from top to bottom corresponding to $Q = 0.01, 0.1, 1, 10$, and $100$, respectively are the bounds from various DM direct detection experiments such as Dark-Side-50 \cite{DarkSide:2018bpj}, PandaX-II \cite{PandaX-II:2017hlx}, LUX \cite{LUX:2019npm}, and XENON1T \cite{XENON:2020gfr}. 
The regions excluded by 
BABAR \cite{Lees:2014xha}, LHCb dark photon searches~\cite{Ilten:2015hya, Ilten:2016tkc,Aaij:2019bvg}, 
by various beam dump experiments, and COHERENT-CsI data~\cite{Akimov:2017ade} are depicted by the light-green, light-orange, cyan, and gray shaded region, respectively (see text for details). 
The light-green region \cite{Bionta:1987qt, Dent:2012mx, Kazanas:2014mca} is excluded from supernova SN 1987 observation \cite{Kamiokande-II:1987idp, Hong:2020bxo, Shin:2021bvz}. 
The search reach of the various ``Lifetime Frontier" experiments FASER~\cite{FASER:2018eoc}, planned FASER 2, Belle II~\cite{Dolan:2017osp}, LDMX~\cite{Berlin:2018bsc}, and SHiP~\cite{Alekhin:2015byh} experiments are depicted by the dark-pink region, light-pink region, purple lines, and orange dashed line, respectively. 
}
\label{fig:FO}
\end{figure}

In Fig.~\ref{fig:FO}, we plot $g_{BL}$ as a function of $m_{Z^\prime}$ for the five benchmark $Q$ values. 
The diagonal blue lines from top to bottom correspond to $Q = 0.01, 0.1, 1, 10$ and $ 100$. Along these lines, the desired relic density $\Omega_{DM} h^2 = 0.120$ is satisfied while regions to the right of each of these lines are excluded because they lead to an overabundance of DM.  
It shows that the gauge coupling values roughly scale as $1/Q$ for fixed $m_{Z^\prime}$.  
This behavior can be roughly understood by noting that for a fixed $m_{Z^\prime}$, $\langle \sigma v\rangle \propto (Q g_{BL})^2 \simeq 1$ pb is necessary to reproduce the observed DM abundance. 
The diagonal dashed lines are very well fitted by 
\bea
g_{BL} \times Q \simeq 3.46\times 10^{-5} \left(\frac{m_{Z^\prime}}{10 \; {\rm GeV}}\right). 
\eea
Figure~\ref{fig:FO} also shows that 
the excluded regions from the current $Z^\prime$ boson search experiments and the future search regions by planned/proposed experiments.
%
%
It shows that $Z^\prime$ searches at current (future) experiments are sensitive to small (large) values of DM charge $Q$. 
The details about the various experiments are discussed below.

For DM mass $m_{DM} \gtrsim 2 $ GeV, the direct DM detection experiments provide stringent constraint on the spin-independent DM elastic scattering cross-section $\sigma_{SI}$ with nucleon. 
For $2 \lesssim m_{DM}\lesssim 6 ~ {\rm GeV }$ the most stringent bounds are from Dark-Side-50 \cite{DarkSide:2018bpj}, PandaX-II \cite{PandaX-II:2017hlx} and LUX \cite{LUX:2019npm} experiments, while XENON1T \cite{XENON:2020gfr} gives the most stringent bound for $m_{DM} > 6$ GeV.  
The combined bound from these experiments can be found in  Ref.~\cite{Mohapatra:2019ysk}. 
In our case, the DM-nucleon cross section is given by 
\bea
\sigma_{SI} \simeq \frac{1}{\pi} Q^2 g_{BL}^4 \frac{\mu^2}{m_{Z^\prime}^4}, 
\eea
where $\mu = m_{\chi} m_{N} /( m_{\chi} + m_{N})$ is the reduced mass for the DM-nucleon system with the nucleon mass $m_N = 0.983$ GeV. 
For fixed $Q$ values and using $m_{\chi} \simeq m_{Z^\prime}/2$, together with upper bounds on $\sigma_{SI}$ in Ref.~\cite{Mohapatra:2019ysk}, we evaluate the upper-bound on $g_{BL}$ as a function of $m_{Z^\prime}$.  
In Fig.~\ref{fig:FO}, the bounds for $Q = 0.01, 0.1,$ and $1$ are depicted by the curved black lines from top to bottom for  $m_{Z^\prime} \gtrsim 4$ GeV, which corresponds to $m_\chi \simeq m_{Z^\prime}/2  > 2$ GeV.

Next let us discuss the parameter region excluded by the various $Z^{\prime}$ boson search experiments in Fig.~\ref{fig:FO}. 
The exclusion from dark photon search by electron-positron collider BABAR \cite{Lees:2014xha} is depicted by the light-green region. 
The light-orange region shows the exclusion from LHCb search for $Z^{\prime}$ decaying to $ \mu^{+} \mu^{-}$.  
The cyan color depicts  the exclusion region from dark photon searches at various beam dump experiments. 
These include electron beam dump experiments, namely, E141 \cite{Riordan:1987aw}, E137 \cite{Bjorken:1988as}, E774 \cite{Bross:1989mp}, KEK \cite{Konaka:1986cb}, Orsay \cite{Davier:1989wz}, and NA64 \cite{Banerjee:2019hmi}, as well as proton beam dump experiments, namely, $ \nu $-CAL I \cite{Blumlein:1990ay}, proton bremsstrahlung \cite{Blumlein:2013cua}, CHARM \cite{Bergsma:1985qz}, NOMAD \cite{Astier:2001ck}, and PS191 \cite{Bernardi:1985ny}. 
We have utilized \texttt{Darkcast}~\cite{Ilten:2018crw} to recast the 90\% confidence level bounds from LHCb, BABAR, and beam dump experiments. 
The gray shaded region is from the COHERENT collaboration~\cite{COHERENT:2019kwz}, as analyzed in~\cite{delaVega:2021wpx}, in search for sub-GeV neutrino DM through coherent elastic neutrino-nucleus scattering processes. 
The light-green region \cite{Bionta:1987qt, Dent:2012mx, Kazanas:2014mca} is excluded from supernova SN 1987 observation \cite{Kamiokande-II:1987idp, Hong:2020bxo, Shin:2021bvz}. 
For $m_{Z^\prime} \lesssim100$ MeV, $Z^\prime$ can be excessively produced from $e^+e^-$ and $\nu \nu$ annihilations in the supernova core and escape, resulting in a considerable loss of energy.

The expected reach of various planned/upcoming ``Lifetime Frontier" experiments are also shown in Fig.~\ref{fig:FO}. 
The light-pink region shows the search reach of the recently approved ForwArd Search ExpeRiment (FASER) 
\cite{Feng:2017vli,FASER:2018eoc,FASER:2019aik}, which will be operational alongside LHC Run-3. 
The planned upgrade of FASER 2 expected to be operational at the High-Luminosity LHC should significantly improve the search reach as depicted by the dark-pink region in Fig.~\ref{fig:FO}. 
The expected search reach of future Belle-II~\cite{Dolan:2017osp} experiment is depicted by the dashed purple line, and the orange-dashed line depicts the reach of the proposed LDMX experiment~\cite{Berlin:2018bsc}. 
The experimental reach of \textit{Search for Hidden Particles} (SHiP) \cite{Alekhin:2015byh}, a proposed experiment at CERN, is depicted by the magenta contour.


\subsection{Freeze-in Dark Matter }\label{sec:FreezeIN}
In the \FI dark matter scenario the DM abundance slowly builds up to a constant value after starting from an effectively  zero initial abundance. 
At late times the expansion rate of the Universe dominates and effectively switches off the DM production such that the DM abundance saturates to a constant value.

\begin{figure}[t]
\begin{center}
\includegraphics[scale=0.6]{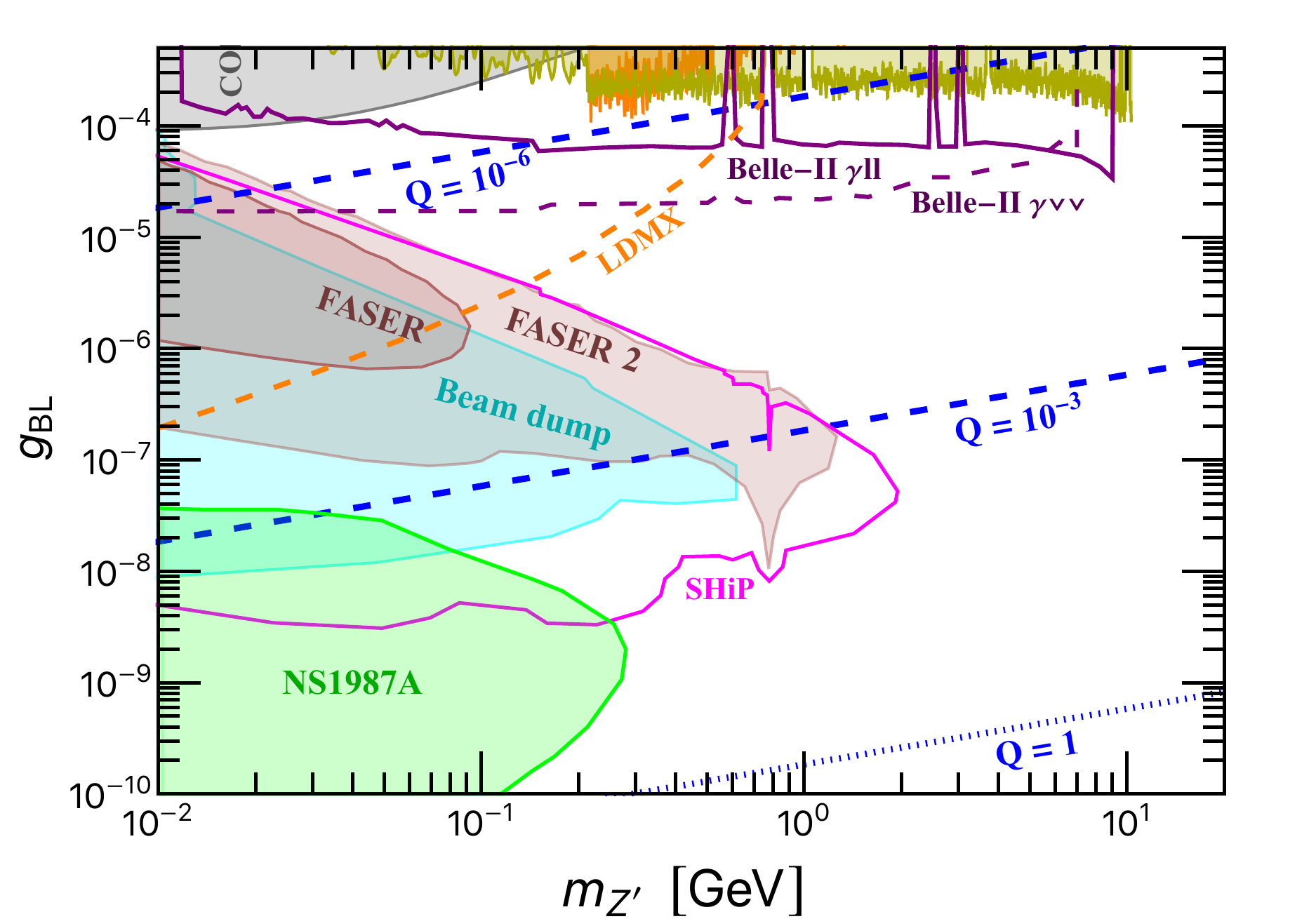}    
\end{center}
\caption{\footnotesize
Same as Fig.~\ref{fig:FO} but for \FI mechanism. 
The diagonal blue lines from top to bottom indicate $Q = 1, 10^{-3}, 10^{-6}$, respectively. 
}
\label{fig:FI}
\end{figure}

For $m_\chi \ll m_{Z^\prime}$, the DM production is effective only for temperature $T \gtrsim m_{Z^\prime}$, namely, $x = m_\chi /T= (m_\chi /m_{Z^\prime}) (m_{Z^\prime}/T) \ll 1$. 
In this case,  $K_2 (x) \simeq 2/x^2$, and hence the thermally averaged cross section in Eq.~(\ref{eq:TAvgCs}) is given by \cite{Okada:2020cue}
\bea
\langle \sigma v\rangle \simeq \frac{\pi }{32} (Q g_{BL})^2\left(\frac{m_{Z^\prime}^3}{m_\chi^5} \right)K_1 \left(\frac{ m_{Z^\prime}}{m_\chi} x \right) x^5 \;.
\label{eq:CSFI}
\eea
Together with the initial condition $Y(x\ll 1) = 0$, we numerically solve the Boltzmann equation in Eq.~(\ref{eq:BE}).
The free parameters  are the DM charge $Q$, $g_{BL}$, $m_{Z^\prime}$, and $m_\chi$. 
In our analysis, we fix $m_\chi = 10$ keV, for example,  and show our result  for $g_{BL}$ as a function of $m_{Z^\prime}$ for various values of $Q$ in Fig.~\ref{fig:FI}. 
The diagonal blue lines from top to bottom  correspond to $Q = 10^{-6}, 10^{-3},$ and $1$, respectively. 
Along these lines, $100\%$ of the observed DM is reproduced while regions to the right of each of these lines are excluded due to the DM overabundance.  
As in the \FO case, for fixed $m_{Z^\prime}$, $g_{BL}$ values roughly scale as $1/Q$. 
However, for same $Q$ values, the $g_{BL}$ values of interest are much smaller compared to the \FO scenario. 
These values are within the search reach of future experiments such as FASER 2 and SHiP.  
Note that the direct DM detection bounds are not applicable for the \FI scenario because the DM ($m_{\chi} = 10 $ keV) is much lighter than the nucleon such that the recoil nucleon energy is too small to be detected.

%

Here we show a rough estimate of the DM relic abundance, which turns out to be a very good approximation to the numerical result.
Our discussion follows Ref.~\cite{Okada:2020cue}. 
Since $K_1 (y) \propto e^{- y}$ and $K_1 (y) \sim 1/y$ for $y \geq 1$ and $y \leq 1$, respectively, Eq.~(\ref{eq:CSFI}) is approximated to be 
\bea
\langle\sigma v_{rel} \rangle \simeq \left(\frac{\pi}{32}\right) \left(g_{BL} Q\right)^2 \left(\frac{m_{Z^\prime}^2}{m_\chi^4}\right) x^4 \;,  
\eea 
for $x\lesssim m_{\chi} /m_{Z^\prime} $, while $\langle\sigma v_{rel} \rangle \simeq 0$ for $x > m_\chi/m_{Z^\prime}$. 
Together with the initial condition $Y(x\ll 1) = 0$, the Boltzmann equation in Eq.~(\ref{eq:BE}) can be solved analytically for the DM yield at late time \cite{Okada:2020cue}:
\bea
Y(\infty) \sim Y(x=m_\chi/m_Z^\prime) \simeq 8.5 \times 10^{-6} \left(g_{BL} Q\right)^2 \left(\frac{M_P}{m_Z^\prime}\right) \;.
\eea 
Substituting this to Eq.~(\ref{eq:relicabundance}) and requiring $\Omega_{DM}h^2 = 0.12$, we obtain a bound on the $B-L$ gauge coupling,
\bea
g_{BL} \times Q \simeq 2.5\times 10^{-12} \left(\frac {m_{Z^\prime}}{m_\chi}\right)^{1/2} \;.
\eea
We find that this result is very close to that obtanined by numerical analysis.

\section{Conclusion }\label{sec:Conclusion}
%
We have studied a $U(1)_{B-L}$ model with $Z^\prime$ portal Dirac fermion dark matter (DM) $\chi$  that weakly couples to  $Z^\prime$. 
The DM $\chi$ carries an arbitrary $B-L$ charge $Q\neq \pm1, \pm 3$  which ensures its stability. 
We have focused on a light $Z^\prime$ with mass $m_{Z^\prime}$ in the range between $10^{-2}$ GeV and a few GeV. 
Since the $Z^\prime$ mediated interactions with the Standard Model (SM) particles determine the relic abundance of $\chi$, the model predicts a one-to-one correspondence between the collider searches for $Z^\prime$ and DM physics.

We have examined both the \FO and \FI DM scenarios. 
The relic abundance of the $\chi$ is determined by $m_{Z^\prime}$, $Q$, $m_\chi$, and the $B-L$ gauge coupling $g_{BL}$. 
In the \FO scenario, the observed DM abundance is reproduced near the $Z^\prime$ resonance, $m_\chi \simeq m_{Z^\prime}/2$. 
In the \FI scenario, we consider $m_\chi \ll m_{Z^\prime}$. 
In both \FO and \FI scenarios, we have shown that for a  fixed value of $m_{Z^\prime}$, the desired $g_{BL}$ values roughly scale as $1/Q$. 
For same $Q$ values, the $g_{BL}$ values of interest in \FI scenario are much smaller compared to the \FO scenario. 

In both \FO and \FI scenarios, for fixed $Q$ values, we have shown that the $g_{BL}$ values necessary to reproduce the observed abundance of $\chi$ is accessible at various current and future planned and/or proposed experiments. 
These include  ongoing experiments including  BABAR, LHCb, COHERENT and various electron and proton beam dump experiments which constrain the allowed parameter space. 
The allowed parameter region can be probed by upcoming ``Lifetime Frontier" experiments such as FASER, FASER 2 (a planned upgrade of FASER) and  Belle-II, and the proposed LDMX and SHiP experiments.

\acknowledgments
\noindent
The work is supported by the Istituto Nazionale di Fisica Nucleare, INFN, through the Theoretical Astroparticle Physics, TAsP, project (N.N.), the United States Department of Energy grant DE-SC0012447 (N.O.), DE-SC0013880 (Q.S. and D.R.) and the M. Hildred Blewett Fellowship of the American Physical Society, www.aps.org (S.O.).

\bibliographystyle{utphys}
\bibliography{references}

\end{document}